\documentclass[prb,preprint,superscriptaddress,longbibliography]{revtex4-1}
\usepackage{graphicx}
\usepackage{amsmath}
\usepackage{amssymb}
\usepackage[usenames,dvipsnames]{color}

\begin{document}

\title{Topological superconductivity in a van der Waals heterostructure} 

\author{Shawulienu Kezilebieke}
\email{Email: kezilebieke.shawulienu@aalto.fi, teemu.ojanen@tuni.fi, peter.liljeroth@aalto.fi}
\affiliation{Department of Applied Physics, Aalto University, FI-00076 Aalto, Finland}

\author{Md Nurul Huda}
\affiliation{Department of Applied Physics, Aalto University, FI-00076 Aalto, Finland}

\author{Viliam Va\v{n}o}
\affiliation{Department of Applied Physics, Aalto University, FI-00076 Aalto, Finland}

\author{Markus Aapro}
\affiliation{Department of Applied Physics, Aalto University, FI-00076 Aalto, Finland}

\author{Somesh C. Ganguli}
\affiliation{Department of Applied Physics, Aalto University, FI-00076 Aalto, Finland}

\author{Orlando J. Silveira}
\affiliation{Department of Applied Physics, Aalto University, FI-00076 Aalto, Finland}

\author{Szczepan G\l odzik}
\affiliation{Institute of Physics, M. Curie-Sk\l odowska University, 20-031 Lublin, Poland}

\author{Adam S. Foster}
\affiliation{Department of Applied Physics, Aalto University, FI-00076 Aalto, Finland}
\affiliation{PI Nano Life Science Institute (WPI-NanoLSI), Kanazawa University, Kakuma-machi, Kanazawa 920-1192, Japan}
	
\author{Teemu Ojanen}
\email{Email: kezilebieke.shawulienu@aalto.fi, teemu.ojanen@tuni.fi, peter.liljeroth@aalto.fi}
\affiliation{Computational Physics Laboratory, Physics Unit, Faculty of Engineering and Natural, Sciences, Tampere University, PO Box 692, FI-33014 Tampere, Finland}
\affiliation{Helsinki Institute of Physics PO Box 64, FI-00014, Finland}

\author{Peter Liljeroth}
\email{Email: kezilebieke.shawulienu@aalto.fi, teemu.ojanen@tuni.fi, peter.liljeroth@aalto.fi}
\affiliation{Department of Applied Physics, Aalto University, FI-00076 Aalto, Finland}

\date{\today}

\maketitle

\textbf{The designer approach has become a new paradigm in accessing novel quantum phases of matter \cite{Geim2013,Novoselov2016_review,Cao2018UnconventionalSuperlattices,Gibertini2019,Yan2019engineered,Khajetoorians2019designer,Lutchyn2018_review}. Moreover, the realization of exotic states such as topological insulators, superconductors and quantum spin liquids often poses challenging or even contradictory demands for any single material\cite{Nayak2008,Sato2017_review,Lutchyn2018_review}. For example, it is presently unclear if topological superconductivity, which has been suggested as a key ingredient for topological quantum computing, exists at all in any naturally occurring material \cite{Mourik2012science,Nadj-Perge2014,Menard2017_NatComm,Kim2018,palacio,Wang2020_Science}. This problem can be circumvented by using designer heterostructures combining different materials, where the desired physics emerges from the engineered interactions between the different components. Here, we employ the designer approach to demonstrate two major breakthroughs -- the fabrication of van der Waals (vdW) heterostructures combining 2D ferromagnetism with superconductivity and the observation of 2D topological superconductivity. We use molecular-beam epitaxy (MBE) to grow two-dimensional islands of ferromagnetic chromium tribromide (CrBr$_3$)\cite{chen2019direct} on superconducting niobium diselenide (NbSe$_2$) and show the signatures of one-dimensional Majorana edge modes using low-temperature scanning tunneling microscopy (STM) and spectroscopy (STS). The fabricated two-di\-men\-sion\-al vdW hetero\-structure provides a high-quality controllable platform that can be integrated in device structures harnessing topological superconductivity. Finally, layered heterostructures can be readily accessed by a large variety of external stimuli potentially allowing external control of 2D topological superconductivity through electrical\cite{Jiang2018}, mechanical\cite{Wu2019}, chemical\cite{Jiang2018a}, or optical means\cite{Zhang2019}.}

There has been a surge of interest in designer materials that would realize electronic responses not found in naturally occurring materials \cite{Geim2013,Novoselov2016_review,Cao2018UnconventionalSuperlattices,Gibertini2019,Yan2019engineered,Khajetoorians2019designer,Lutchyn2018_review}. Topological superconductors are one of the main targets of these efforts and they are currently attracting intense attention due to their potential as building blocks for Majorana-based qubits for topological quantum computation\cite{Nayak2008,Sato2017_review,Lutchyn2018_review}. Majorana zero-energy modes (MZM) have been reported in several different experimental platforms, with the most prominent examples being semiconductor nanowires with strong spin-orbit coupling and ferromagnetic atomic chains proximitized with an s-wave superconductor \cite{Mourik2012science,Nadj-Perge2014,Sato2017_review,Kim2018,Lutchyn2018_review,Liu2019_review,Jaeck2019}. It is also possible to realize MZMs in vortex cores on a proximitized topological insulator surface\cite{Fu2008,Sun2016} or on FeTe$_{0.55}$Se$_{0.45}$ superconductor surface \cite{Zhang2018_Science,Wang2018_Science,Zhu2020_science}. In these cases the MZM were spectroscopically identified as zero energy conductance signals that are localized at the ends of the one dimensional (1D) chain or in the vortex core. The evidence of the Majorana states in these various platforms consists of subgap conductance peaks and the features occurring at system boundaries and defects are consistent with their topological origin. Unfortunately, different disorder-induced states are known to mimic the Majorana conductance signals and the status of the observations remains presently inconclusive. The ultimate proof of the existence of Majorana zero modes would be obtained if their non-Abelian exchange statistics or braiding could be demonstrated\cite{Lutchyn2018_review}.

\begin{figure}[!h]
	\centering
	\includegraphics[width=0.82\textwidth]{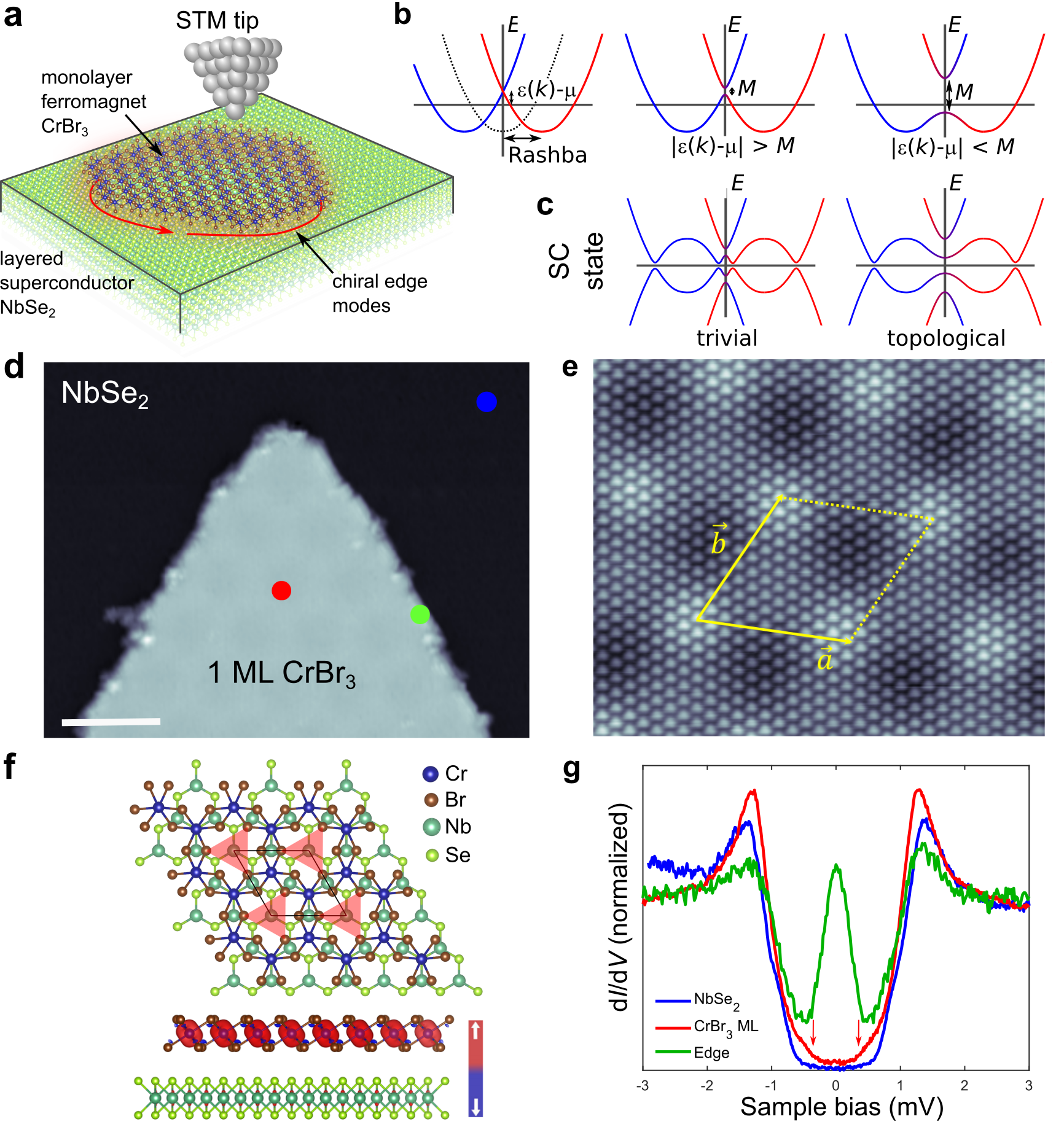}
	\caption{\textbf{Realization of topological superconductivity in CrBr$_3$-NbSe$_2$ heterostructures.} \textbf{a,} Schematic of the experimental setup. \textbf{b,c,} Schematic of the bandstructure engineering to realize topological superconductivity. Effect of adding spin-orbit interactions and weaker and stronger Zeeman-type magnetization on the low-energy band structure in the normal (b) and superconducting states (c). \textbf{d,} STM image of a monolayer thick CrBr$_3$ island grown on NbSe$_2$ using MBE (STM feedback parameters: $V_\mathrm{bias} = +1$ V, $I$ = 10 pA, scale bar: 10 nm). \textbf{e,} Atomically resolved image on the CrBr$_3$ layer (STM feedback parameters: $V_\mathrm{bias} = +1.7$ V, $I = 0.5$ nA, image size: $19 \times 19$ nm$^2$). \textbf{f,} Calculated structure and the induced spin-polarization from density-functional theory calculations. \textbf{g,} Experimental d$I$/d$V$ spectroscopy on the NbSe$_2$ substrate (blue), the middle of the CrBr$_3$ island (red) and at the edge of the CrBr$_3$ island (green) measured at $T=350$ mK. }
	\label{fig1}
\end{figure}

In two-dimensional systems, 1D dispersive chiral Majorana fermions are expected to localize near the edge of the system (Fig.~\ref{fig1}a). For example, it was proposed that the dispersing Majorana states can be created at the edges of an island of magnetic adatoms on the surface of an s-wave superconductor\cite{Roentynen2015,Li2016,Rachel2017}. Experimentally, promising signatures of such 1D chiral Majorana modes have recently been reported around nanoscale magnetic islands either buried below a single atomic layer of Pb\cite{Menard2017_NatComm}, or adsorbed on a Re substrate\cite{palacio}, and in domain walls in FeTe$_{0.55}$Se$_{0.45}$ \cite{Wang2020_Science}. However, these types of systems can be sensitive to disorder and may require interface engineering through, \emph{e.g.}, the use of an atomically thin separation layer. In addition, it is difficult to incorporate these materials into device structures. These problems can be circumvented in van der Waals (vdW) heterostructures, where the different layers interact only through vdW forces\cite{Geim2013}. VdW heterostructures naturally allow for very high quality interfaces and a multitude of practical devices have been demonstrated. While vdW materials with a wide range of properties have been discovered, ferromagnetism has been notably absent until recent discoveries of atomically thin Cr$_2$Ge$_2$Te$_6$\cite{Gong2017_Cr2Ge2Te6}, CrI$_3$\cite{Huang2017_CrI3} and CrBr$_3$\cite{Ghazaryan2018,Zhang2019}. The first reports relied on mechanical exfoliation for the sample preparation, but CrBr$_3$\cite{chen2019direct} and Fe$_3$GeTe$_2$\cite{Liu2017_Fe3Gete2} have also been grown using molecular-beam epitaxy (MBE) in ultra-high vacuum (UHV). This is essential for realizing clean edges and interfaces.

The recently discovered monolayer ferromagnet transition metal trihalides combined with transition metal dichalcogenide (TMD) superconductors form an ideal platform for realizing 2D topological superconductivity (Fig.~\ref{fig1}a). Here, we use MBE to grow high-quality monolayer ferromagnet CrBr$_3$ on a NbSe$_2$ superconducting substrate. The mirror symmetry is broken at the interface between the different materials and this lifts the spin degeneracy due to the Rashba effect. Therefore, we have all the necessary ingredients -- magnetism, superconductivity and Rashba spin-orbit coupling -- required to realize a designer topological superconductor \cite{Sato2009_PRB,Sau2010_PRL}. The use of a van der Waals heterostructure has significant advantages: They can potentially be manufactured by simple mechanical exfoliation, the interfaces are naturally very uniform and of high quality, and the structures can be straightforwardly integrated in device structures. Finally, layered heterostructures can be readily accessed by a large variety of external stimuli making external control of 2D topological superconductivity potentially possible by electrical\cite{Jiang2018}, mechanical\cite{Wu2019}, chemical\cite{Jiang2018a}, and optical approaches\cite{Zhang2019}. 

Pioneering theoretical works \cite{Sato2009_PRB,Sau2010_PRL} demonstrated that topological superconductivity may arise from a combination of out-of-plane ferromagnetism, superconductivity and Rashba-type spin-orbit coupling, as illustrated in Fig.~\ref{fig1}b,c. In this scheme, the Rashba coupling lifts the spin-degeneracy of the conduction band while Zeeman splitting due to proximity magnetization lifts the remaining Kramers degeneracy. Adding superconductivity creates a particle-hole symmetric band structure and the superconducting pairing opens gaps at the Fermi energy. In our theoretical model for magnetically covered NbSe$_2$, a similar picture arises for the real band structure around any of the high symmetry points of the hexagonal Brillouin zone ($\Gamma$, $\mathrm{K}$, or $\mathrm{M}$) where Rashba coupling vanishes. Depending on the magnitude of the magnetization induced gap $M$ and the position of the Fermi energy $\mu$, the system enters a topological phase when $|\epsilon(\Vec{k_0})-\mu|\leq M$, where $\epsilon(\Vec{k_0})$ is the energy of the band crossing at the high symmetry point in the absence of magnetization. This is due to the created effective $p$-wave pairing symmetry. 

In  Fig.~\ref{fig1}d, we show a constant-current scanning tunneling microscopy (STM) image of the CrBr$_3$ island grown on a freshly cleaved bulk NbSe$_2$ substrate by MBE (see Methods for details). The CrBr$_3$ islands show a well-ordered moir\'e superstructure with 6.3 nm periodicity arising from the lattice mismatch between the CrBr$_3$ and the NbSe$_2$ layers. Fig.~\ref{fig1}e shows an atomically resolved STM image of the CrBr$_3$ monolayer, revealing periodically spaced triangular protrusions. These features are formed by the three neighbouring Br atoms as highlighted in the Fig.~\ref{fig1}f (red triangle) showing the fully relaxed geometry of CrBr$_3$/NbSe$_2$ heterostructure obtained through density functional theory (DFT) calculations (see Methods for details). The measured in-plane lattice constant is 6.5 \AA, consistent with the recent experimental value (6.3 \AA) of monolayer CrBr$_3$ grown on graphite \cite{chen2019direct} and our DFT calculations. DFT calculations further confirm that the CrBr$_3$ monolayer retains its ferromagnetic ordering with a magnetocrystalline anisotropy favouring an out-of-plane spin orientation as shown in Fig.~\ref{fig1}f. We confirmed the ferromagnetism of the CrBr$_3$ islands on NbSe$_2$ experimentally with magneto-optical Kerr effect measurements. The magnetization density (Fig.~\ref{fig1}f) shows that the magnetism arises from the partially filled $d$ orbitals of the Cr$^{3+}$ ion. While the largest magnetization density is found close to the Cr atoms, there is also significant proximity induced magnetization on the Nb atoms in the underlying NbSe$_2$ layer. 

We probe the emerging topological superconductor phase with scanning tunneling spectroscopy (STS) measurements at a temperature of $T=350$ mK. Fig.~\ref{fig1}g shows experimental d$I$/d$V$ spectra (raw data) taken at different locations indicated in Fig.~\ref{fig1}d (marked by filled circles). The d$I$/d$V$ spectrum of bare NbSe$_2$ has a hard gap with an extended region of zero differential conductance around zero bias, which can be fitted by the McMillan two-band model\mbox{\cite{Noat2015_nbse2}} (see Extended Data Fig.~\mbox{\ref{ext-fig:fitting}}a). In contrast, in the spectra taken in the middle of the CrBr$_3$ island, we observe pairs of conductance onsets at $\pm 0.3$ mV around zero bias (red arrows). The magnetization causes the formation of energy bands (dubbed Shiba bands) that exist inside the superconducting gap of the substrate \cite{Sato2009_PRB,Sato2017_review}. Spin-orbit interactions can drive the system into a topological phase with associated closing and reopening of the gap between the Shiba bands.

We observe edge modes consistent with the expected Majorana modes along the edge of the magnetic island that are the hallmark of 2D topological superconductivity \cite{Sato2017_review,Menard2017_NatComm,palacio}. The spectroscopic feature of the Majorana edge mode appears inside the gap defined by the Shiba bands (the topological gap) and is centred around the Fermi level ($E_\mathrm{F}$). This is a further indication that the edge states are indeed topological edge modes. A typical spectrum taken at the edge of the CrBr$_3$ island is shown in Fig.~\ref{fig1}g, where a peak localized at $E_\mathrm{F}$ is clearly seen together with side features stemming from the Shiba bands (detailed analysis shown in Extended Data Fig.~\mbox{\ref{ext-fig:fitting}}c-f).

\begin{figure}[!h]
	\centering
	\includegraphics[width=0.9\textwidth]{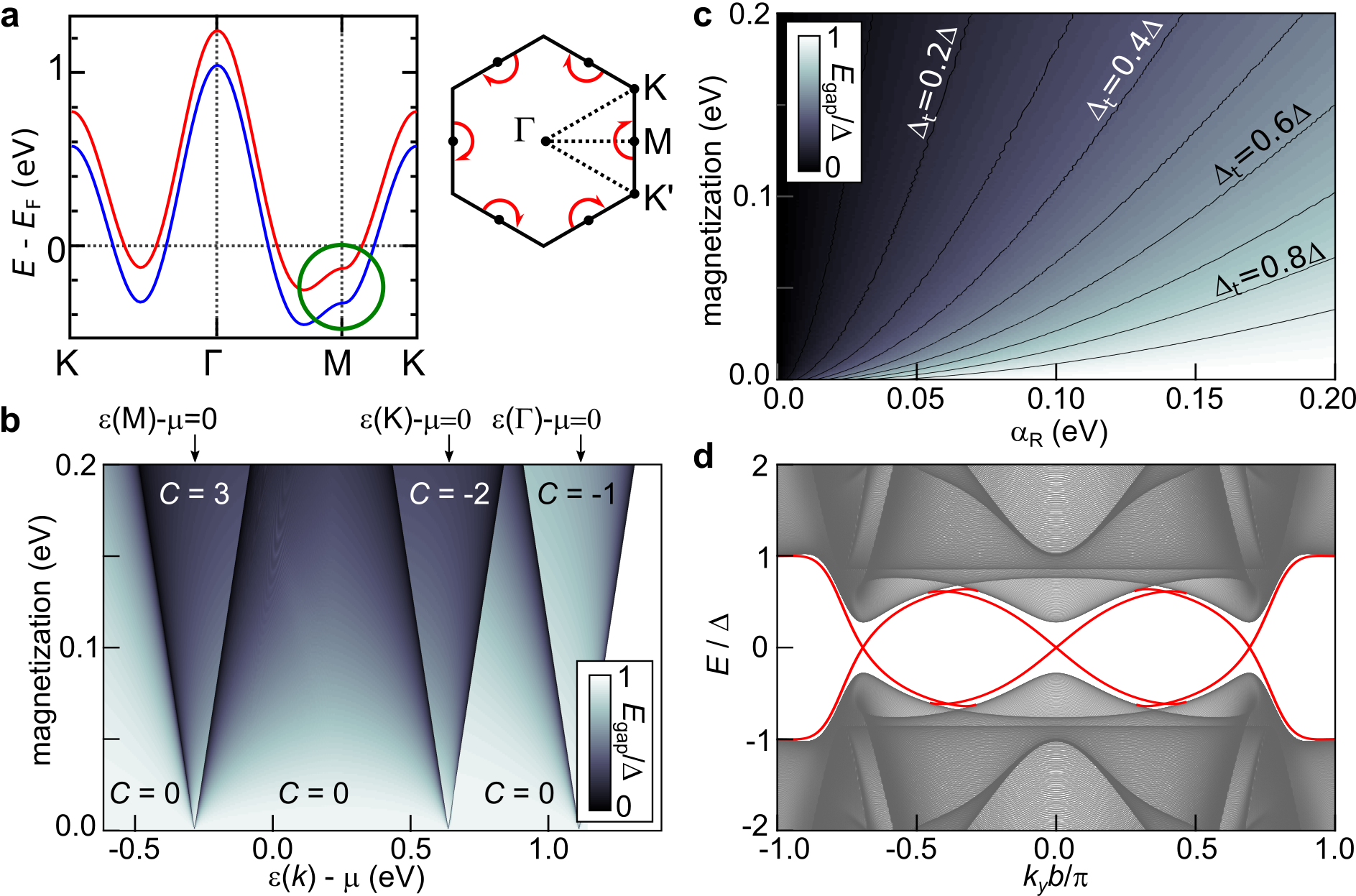}
	\caption{\textbf{Electronic structure of CrBr$_3$-NbSe$_2$ heterostructures.} \textbf{a,} The band structure of the spin-split Nb $d$-band used in the effective model for topological superconductivity with magnetization $M=100$ meV. The inset shows the 1$^\mathrm{st}$ Brillouin zone, where the six M-points and the Rashba texture around them has been highlighted. \textbf{b,} Calculated phase diagram of the magnetized NbSe$_2$ based on the effective low-energy model. The color scale indicates the energy gap $E_\mathrm{gap}$ (in the units of $\Delta$). \textbf{c,} The calculated topological gap $\Delta_t$ as a function of the Rashba and magnetization energies (in units of the superconducting gap $\Delta$). \textbf{d,} Calculated band structure of the topological phase based on a phenomenological tight-binding model (see SI for details).}
	\label{fig2}
\end{figure}

We develop an effective low-energy model (see Supplementary Information (SI) for details) based on earlier work on individual magnetic impurities on a bulk NbSe$_2$ substrate \mbox{\cite{Menard2015_NatPhys}} and DFT calculations. The band structure of the Nb $d$-states derived band used in the effective model is shown in Fig.~\ref{fig2}a (direct comparison with DFT is shown in the SI). Topological superconductivity can be generated when magnetization is sufficiently strong to push one of the spin-degenerate bands at a high-symmetry point above the Fermi energy. We identify the observed topological phase as a  state arising from the gap-closing transition at M point with a Chern number $C=3$. While the M-point is at $\sim270$ meV below the Fermi level in our TB model (calculated phase diagram shown in Fig.~\mbox{\ref{fig2}}b), it is estimated to be $\sim 100$ meV below Fermi level in bulk NbSe$_2$ \mbox{\cite{PhysRevLett.102.166402,Ugeda2016NatPhys}}. This implies that for a reasonable magnetization of $M\lesssim100$ meV, we need slight doping to bring the experimental system into the $C=3$ state. This is precisely what we observe by following the Nb d-bands from bare NbSe$_2$ onto a CrBr$_3$ island. There is an upward shift of the NbSe$_2$ bands of $\sim80$ meV under CrBr$_3$ (Extended Data Fig.~\mbox{\ref{ext-fig:shift}}). The two other nontrivial phases that originate from gap closings at the $\Gamma$ point and $K$ points give rise to  topological phases with $C=-1$ and $C=-2$. Realization of either of these phases would require notably larger shifts in chemical potential ($\sim 0.6$ eV), making them improbable for the experimental observations. The absolute values of the nontrivial Chern numbers can be understood by a three-fold rotational symmetry (see SI).
 
The key quantity characterizing robustness of the nontrivial phase is the topological energy gap $\Delta_t$. This scale should be much larger than temperature for the state to be observable in experiment. In the simple parabolic band model this quantity can be estimated by $\Delta_t=\alpha k_F/[(\alpha k_F)^2+M^2]^{1/2}$, where $\alpha$ is the Rashba coupling and $k_F$ the Fermi wavelength \cite{Alicea2010_PRB}. The calculated gap based on our more realistic tight-binding (TB) model is shown in Fig.~\ref{fig2}c. Based on the experimental results shown in Fig.~\ref{fig1}g, the topological gap is $\Delta_t\approx0.3\Delta$. The calculated band structure in a strip geometry corresponding to the experimental gap is shown in Fig.~\ref{fig2}d, where we see the Majorana edge modes crossing the topological gap. The edge modes are seen to coexist with the bulk states in a finite subgap energy window in agreement with experimental observations.

\begin{figure}[!b]
	\centering
	\includegraphics[width=0.95\textwidth]{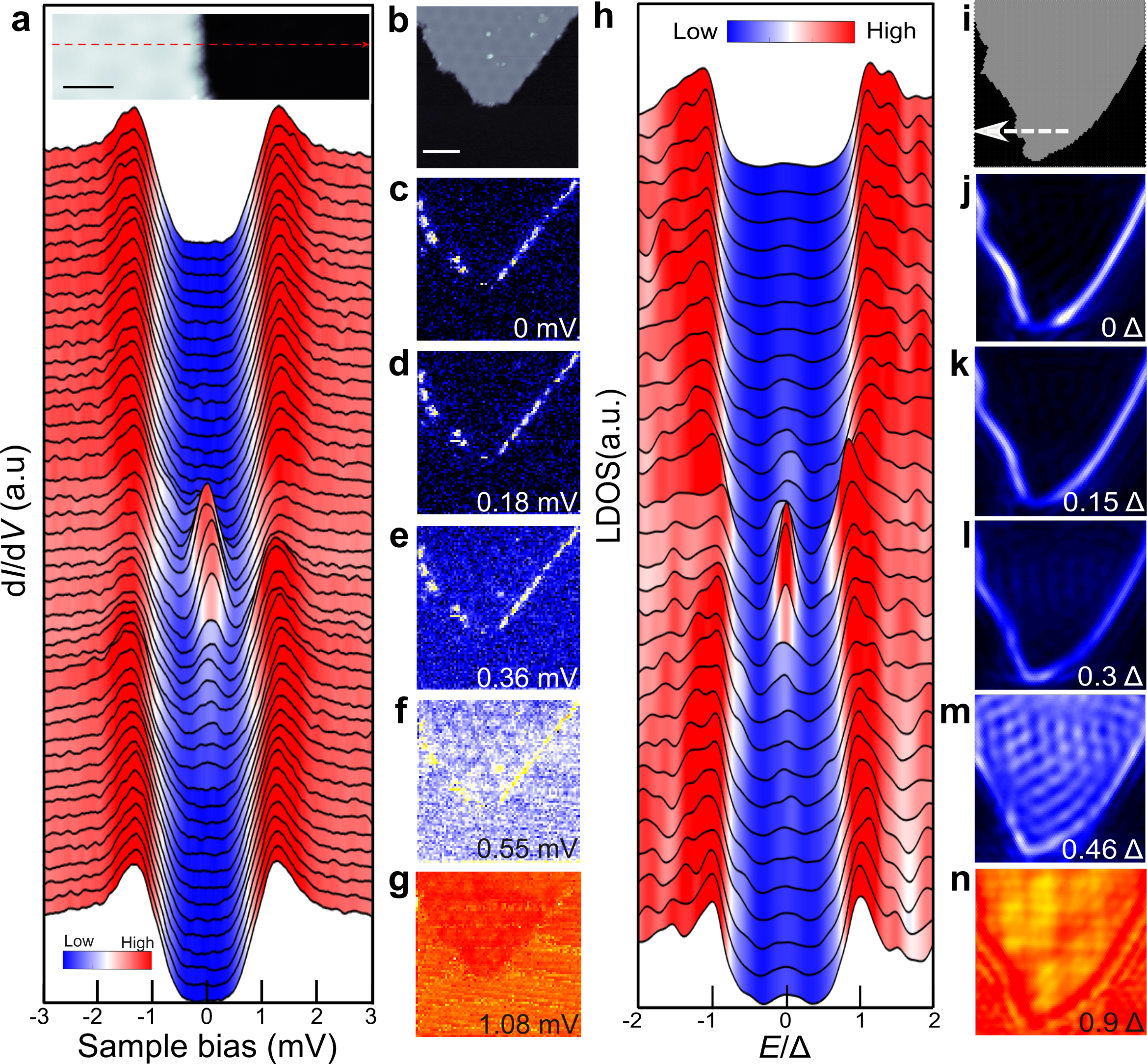}
	\caption{\textbf{Spatially resolved spectroscopy of the Majorana zero modes.} \textbf{a,} d$I$/d$V$ spectroscopy across the edge of the CrBr$_3$ island (STM topography shown on the top). \textbf{b-g,} STM topography and spatially resolved d$I$/d$V$ maps extracted from grid spectroscopy experiments. STM feedback parameters: (a) $V_\mathrm{bias} = +1$ V, $I = 10$ pA; (b) $V_\mathrm{bias} = +0.8$ V, $I = 10$ pA. Scale bars: (a) 4 nm; (b) 12 nm. \textbf{h-n,} Corresponding calculated LDOS across the edge (h) and LDOS maps (j-n) with the island shape shown in (i) (see SI for details). }
	\label{fig3}
\end{figure}

We have carried out spatially resolved d$I$/d$V$ spectroscopy across the edge of the CrBr$_3$ island (Fig.~\ref{fig3}a). The energy dependence of the main feature of the edge mode LDOS is such that it splits off from the top edge of the topological gap inside the CrBr$_3$ island, smoothly crosses the topological gap and merges with its lower edge outside the CrBr$_3$ island. We have recorded grid d$I$/d$V$ spectroscopy maps (Fig.~\ref{fig3}b-g) to probe the spatial evolution of the edge modes. At $E_\mathrm{F}$, the edge modes are confined within $\sim2.4$ nm of the edge of the island (see Extended Data Fig.~\ref{ext-fig:edge-decay}). In addition to the edge mode signature close to the Fermi level, there is also enhanced LDOS at the energies above the topological gap (Fig.~\ref{fig3}e,f) where we also see significant excitations inside the magnetic island. This implies that the edge modes coexist with Shiba bands at energies higher than the topological gap. The theoretically computed LDOS (Fig.~\ref{fig3}h-n, see SI for details) reproduces the essential features of the experimental results. In addition to the universal signatures of topological superconductivity, our theoretical model reproduces a number of experimentally observed system-specific characteristics including 1) the correct edge mode penetration depth which is orders of magnitude smaller than simple estimates (similar to the case of 1D Fe wires on Pb \mbox{\cite{Nadj-Perge2014,peng2015}}), 2) the specific form of the subgap local density of states, which depends on system-specific dispersion of the topological edge modes, and 3) a non-generic coexistence of the topological edge modes and bulk states in a substantial energy window. Finally, we confirm the experimental finding that the distribution of the spectral weight of the edge mode along the edge is non-uniform. This stems from the geometric irregularities of the island boundary with characteristic length scale that is comparable to the edge mode penetration depth. However, the edge modes are not discontinuous along the edge, interference effects near edge irregularities suppress the visibility of the edge mode due to finite experimental resolution.

The experimentally observed edge modes could possess a topologically trivial origin. However, in addition to the near quantitative match with the theoretical results incorporating the main ingredients of the experimental system, the edge mode signature is experimentally very robust. We consistently observe it in our hybrid vdW heterostructures on all CrBr$_3$ islands, irrespective of their specific size and shape (Extended Data Fig.~\ref{ext-fig:examples}). Based on theoretical considerations, to observe the modes for all subgap energies, the linear dimension of the islands should be much larger than the edge mode penetration depth. This condition is clearly fulfilled in our experimental results. To prove the observed edge modes of the hybrid heterostructures are strongly linked to the superconductivity of the NbSe$_2$ substrate, we have carried out experiments in magnetic fields up to 4 T, suppressing superconductivity in the NbSe$_2$ substrate. All features associated with the gap at the centre of the island and the edge modes disappear in the absence of superconductivity in NbSe$_2$ (Extended Data Fig.~\ref{ext-fig:mag-field}). This rules out trivial edge modes as the cause of the observed results. Another non-topological reason for resonances close to the Fermi energy, the Kondo effect, should also be present in the normal state and can hence be ruled out as well. 

In conclusion, our work constitutes two breakthroughs in designer quantum materials. By fabricating vdW heterostructures with 2D ferromagnet epitaxially coupled to superconducting NbSe$_2$, we obtained a near ideal designer structure exhibiting two competing electronic orders. The induced magnetization and spin-orbit coupling renders the superconductor topologically nontrivial, supporting Majorana edge channels which we characterized by STM and STS measurements. The demonstrated heterostructure provides a high-quality platform for electrical devices employing topological superconductivity. 
This is an essential step towards practical devices employing topological superconductivity and the demonstrated system would in principle allow electrical control of the topological phase through electrostatic tuning of the chemical potential.  

\bibliographystyle{naturemag1}
\bibliography{bibliography}

\section*{Acknowledgments}
This research made use of the Aalto Nanomicroscopy Center (Aalto NMC) facilities and was supported by the European Research Council (ERC-2017-AdG no.~788185 ``Artificial Designer Materials''), Academy of Finland (Academy professor funding no.~318995 and 320555, and Academy postdoctoral researcher no.~309975), and the Aalto University Centre for Quantum Engineering (Aalto CQE). SG acknowledges the support of National Science Centre (NCN, Poland) under grant 2017/27/N/ST3/01762. Computing resources from the Aalto Science-IT project and CSC, Helsinki are gratefully acknowledged. ASF has been supported by the World Premier International Research Center Initiative (WPI), MEXT, Japan. 

\section*{Author statements}

\subsection*{Contributions}
SK, TO and PL conceived the experiment, SK, MNH, MA, and SCG carried out the sample growth. SK did the low-temperature STM experiments. SK and VV analyzed the STM data. OJS and ASF planned and carried out the DFT calculations. TO and SG developed the theoretical model and established its implications. SG carried out the numerical calculations. SK, TO and PL wrote the manuscript with input from all coauthors.

\subsection*{Corresponding authors}
Correspondence and requests for materials should be addressed to S.K., T.O., or P.L.

\subsection*{Competing interests}
The authors declare no competing interests.

\newpage

\section*{Methods}
\subsection*{Experimental} 
The CrBr$_3$ thin film was grown on a freshly cleaved NbSe$_2$ substrate by compound source molecular beam epitaxy (MBE). The anhydrous CrBr$_3$ powder of 99 \% purity was evaporated from a Knudsen cell. Before growth, the cell was degassed up to the growth temperature $350^\circ$C until the vacuum was better than $1\times10^{-8}$ mbar. The growth speed was determined by checking the coverage of the as-grown samples by scanning tunneling microscopy (STM). The optimal substrate temperature for the growth of CrBr$_3$ monolayer films was $\sim270^\circ$C. 

After the sample preparation, it was inserted into the low-temperature STM (Unisoku USM-1300) housed in the same UHV system and all subsequent experiments were performed at $T = 350$ mK. STM images were taken in the constant-current mode. d$I$/d$V$ spectra were recorded by standard lock-in detection while sweeping the sample bias in an open feedback loop configuration, with a peak-to-peak bias modulation of 30-50 $\mu$V at a frequency of 707 Hz. Spectra from grid spectroscopy experiments were normalized by the normal state conductance, i.e.~d$I$/d$V$ at a bias voltage corresponding to a few times the superconducting gap.

\subsection*{Density Functional Theory (DFT) calculations}

Calculations were performed with the DFT methodology as implemented in the periodic plane-wave basis VASP code \cite{PhysRevB.54.11169,KRESSE199615}. Atomic positions and lattice parameters were obtained by fully relaxing all structures using the spin-polarized Perdew-Burke-Ernzehof (PBE) functional \cite{PhysRevLett.77.3865} including Grimme's semiempirical DFT-D3 scheme for dispersion correction \cite{doi:10.1063/1.3382344}, which is important to describe the van der Waals (vdW) interactions between the CrBr$_3$ and the NbSe$_2$ layers. The interactions between electrons and ions were described by PAW pseudopotentials, where 4s and 4p shells were added explicitly as semicore states for Nb and 3p shells for Cr. An energy cutoff of 550 eV is used to expand the wave functions and a systematic $k$-point convergence was checked for all structures, with sampling chosen according to each system size. For all systems the total energy was converged to the order of 10$^{-4}$ eV. The convergence criterion of self-consistent field (SCF) computation was set to 10$^{-5}$ eV and the threshold for the largest force acting on the atoms was set to less than 0.012 eV/\AA{}. A vacuum layer of 12 \AA{} was added to avoid mirror interactions between periodic images. Spin polarization was considered in all calculations, where we set an initial out of plane magnetization of 3 $\mu_B$ per Cr atom and 0 otherwise. Band structures were calculated with and without spin-orbit coupling (SOC) effects, and a band unfolding procedure was performed when necessary using the BandUP code \cite{PhysRevB.89.041407, PhysRevB.91.041116}. 

In Extended Data Fig.~\ref{geos} are shown the top and side view of the CrBr$_3$ and NbSe$_2$ monolayers, where the optimized lattice parameters are 6.370 \AA{} and 3.455 \AA{}, respectively. In order to model the CrBr$_3$-NbSe$_2$ heterostructures shown in Figure 1 of the main text, the CrBr$_3$ is vertically stacked on a $2\times2$ supercell of NbSe$_2$. As there is a mismatch of 8.5 \% between their lattice parameters, the CrBr$_3$ was rescaled prior to the full optimization and three different stackings were considered, namely: htCrSe, htCrNbSe and htCrNb (Extended Data Fig.~\ref{geos}c-e, respectively). In htCrSe, one Cr atom is located on top of a Se$_2$ pair, while the other Cr atom is on top of the hollow site of the NbSe$_2$. Similarly, the htCrNb has one Cr on top of the Nb and the other Cr is on top of the hollow site. For the htCrNbSe, one Cr is located on top of the Se$_2$ pair, while the other Cr is on top of the Nb. As can be seen from the data in Table \ref{tabledft}, the fully relaxed lattice parameters $L$ of the heterostructures reveal that the CrBr$_3$ is strained by about 7 \%, while the NbSe$_2$ is compressed by less than 2 \%, which are small enough to not significantly affect their electronic and magnetic properties \cite{PhysRevB98144411, C4NR01486C}. The binding energies $E_b$ in Table \ref{tabledft} reveal that htCrSe is the most energetically stable. For htCrse and htCrNbse, the energy differences between the stacking configurations are relatively small (3.4 meV), and their layer-layer distances $d$ are also comparable. htCrNb is the stacking configuration with highest layer-layer distance among the three stackings considered, resulting in a higher binding energy.

The band structures of the isolated monolayers of CrBr$_3$ and NbSe$_2$ are in the left and middle panels of Extended Data Fig.~\ref{bands-ht}, respectively, while the right panel shows the band structure of htCrSe.  CrBr$_3$ has an out of plane magnetization of 6.0 $\mu_B$, and is a semiconductor with an indirect band gap of 1.39 eV for spin up and 2.63 eV for spin down channels. On the other hand, NbSe$_2$ has a partially filled isolated and spin degenerated band called $d$-band due to the large contributions coming from the Nb $d$-orbitals \cite{PhysRevB.96.155439}. The middle panel of Extended Data Fig.~\ref{bands-ht} shows both the band structure of the $2\times2$ NbSe$_2$ isolated monolayer, which is the supercell used for the CrBr$_3$-NbSe$_2$ heterostructures, and the band structure of the NbSe$_2$ primitive cell in the inset. Considering the band structure of the htCrSe structure in the right panel of Figure \ref{bands-ht}, apart from a bandgap reduction of the CrBr$_3$ spin polarized band structure (which is attributed to its strained configuration \cite{PhysRevB98144411}), the bands of the $2\times2$ NbSe$_2$ are well preserved within the band gaps of the CrBr$_3$. However, a small spin splitting is observed in these in-gap states due to an induced magnetization on the NbSe$_2$ layer. Similar to previous works on CrBr$_3$-TMD heterostructures \cite{PhysRevB.101.205404, PhysRevB.100.085128}, our results show that the magnetization on htCrSe2 (6.097 $\mu_B$) is slightly larger than the magnetization of CrBr$_3$ (6.000 $\mu_B$). 

From now on we discuss in detail the spin splitting of the in-gap bands from the NbSe$_2$ shown in Extended Data Fig.~\ref{unfold}a. Extended Data Fig.~\ref{unfold}b shows the comparison between the bands of the pristine $2\times2$ NbSe$_2$ layer and the bands of htCrSe, both around the $\Gamma$ point, where two sets of spin polarized bands are clearly seen, with spin splittings of 29 and 7 meV. The two sets of bands are obtained because each Nb atom is coupled differently with the CrBr$_3$ layer in the heterostructure. The unfolded bands in Extended Data Fig.~\ref{unfold}c show that, although the overall effect of the magnetization is to push the spin down band above the spin up band, similar to others CrBr$_3$-TMD heterostructures \cite{PhysRevB.101.205404, PhysRevB.100.085128}. There is a discontinuity at the $M$ point as the Nb atoms are no longer identical in htCrSe (note that the $\Gamma$ point at the Brillouin zone (BZ) of $2\times 2$ NbSe$_2$ is equivalent to the $\Gamma$ and $M$ points at the NbSe$_2$ primitive cell's BZ). As the bands of htCrSe around the Fermi level have a major contribution from the $d$ electrons of the Nb atoms, spin-orbit coupling (SOC) effects can drastically affect their dispersion. In Extended Data Fig.~\ref{unfold}d is shown the band structure of htCrSe with SOC. Extended Data Fig.~\ref{unfold}e shows the comparison between the bands of a pristine $2\times2$ NbSe$_2$ layer and the bands of htCrSe, both around the $\Gamma$ point and also considering SOC. There is no SOC splitting along $\Gamma$-$M$ while a large splitting is obtained along the $K$-$\Gamma$ line in the pristine NbSe$_2$ case \cite{PhysRevB.96.155439}, pushing the spin up bands above the spin down bands. For htCrSe, the spin splitting due to SOC is also observed along the $K$-$\Gamma$ line, while the bands along the $\Gamma$-$M$ are spin-polarized due to the induced magnetization. The unfolded bands shown in Extended Data Fig.~\ref{unfold}f reveal that indeed the bands along the $\Gamma$-$M$ are spin polarized due to the induced magnetization (although a small reduction of the splitting is observed when SOC is considered), with the spin down band above the spin up band, while the bands along the $M$-$K$ line have an inverted splitting induced by SOC.

The band structures shown in Extended Data Fig.~\ref{unfold}d,e were obtained considering three different effects simultaneously: induced magnetization + SOC + different coupling between each Nb atom with the CrBr$_3$ layer. Mainly due to the non-equivalency between the Nb atoms, many states are observed crossing the $\Gamma$ point in the band structure of htCrSe, as an extra splitting is observed in the spin polarized bands. However, we stress here that the magnetic band gap at the $M$ point is kept when SOC is taken into account (see Extended Data Fig.~\ref{unfold}f), and the spin inversion observed in the vicinity of the $M$ point in the unfolded band structure is compatible with the spin inversion in the Rashba effect \cite{natureRashba1, natureRashba2, natureRashba3, PhysRevB.92.121403}. However, a precise estimate of the Rashba effect would require the inclusion of SOC effects in a large supercell structure where all Nb atoms couple equally to the CrBr$_3$ layer, or at least large enough to average the coupling from each Nb atom, which is beyond any computational capability. However, judging by the sizeable spin polarized band gap at the $M$ point when SOC is considered and the observed spin inversion, we conclude that the Rashba effect is of the same order of magnitude as the spin splitting due to the induced magnetization. 

Note that our DFT calculations show that all states between the VBM and CBM of the CrBr$_3$ come from the metallic substrate, which is expected for such a weakly interacting system. Estimates of the electron and hole barrier heights based on our DFT calculations \cite{shimada_work_1994,liu_van_2016,Zhang2019} indicate that there is no spontaneous charge transfer either from or to the metallic substrate.

\section*{Data availability statement}
All the data supporting the findings are available from the corresponding authors upon request.

\newpage

\renewcommand{\figurename}{\textbf{Extended Data Figure}}
\setcounter{figure}{0}

\renewcommand{\tablename}{\textbf{Extended Data Table}}
\setcounter{table}{0}

\section*{Extended data figure and table legends}

\begin{figure}[!h]
    \centering
    \includegraphics[width=1\textwidth]{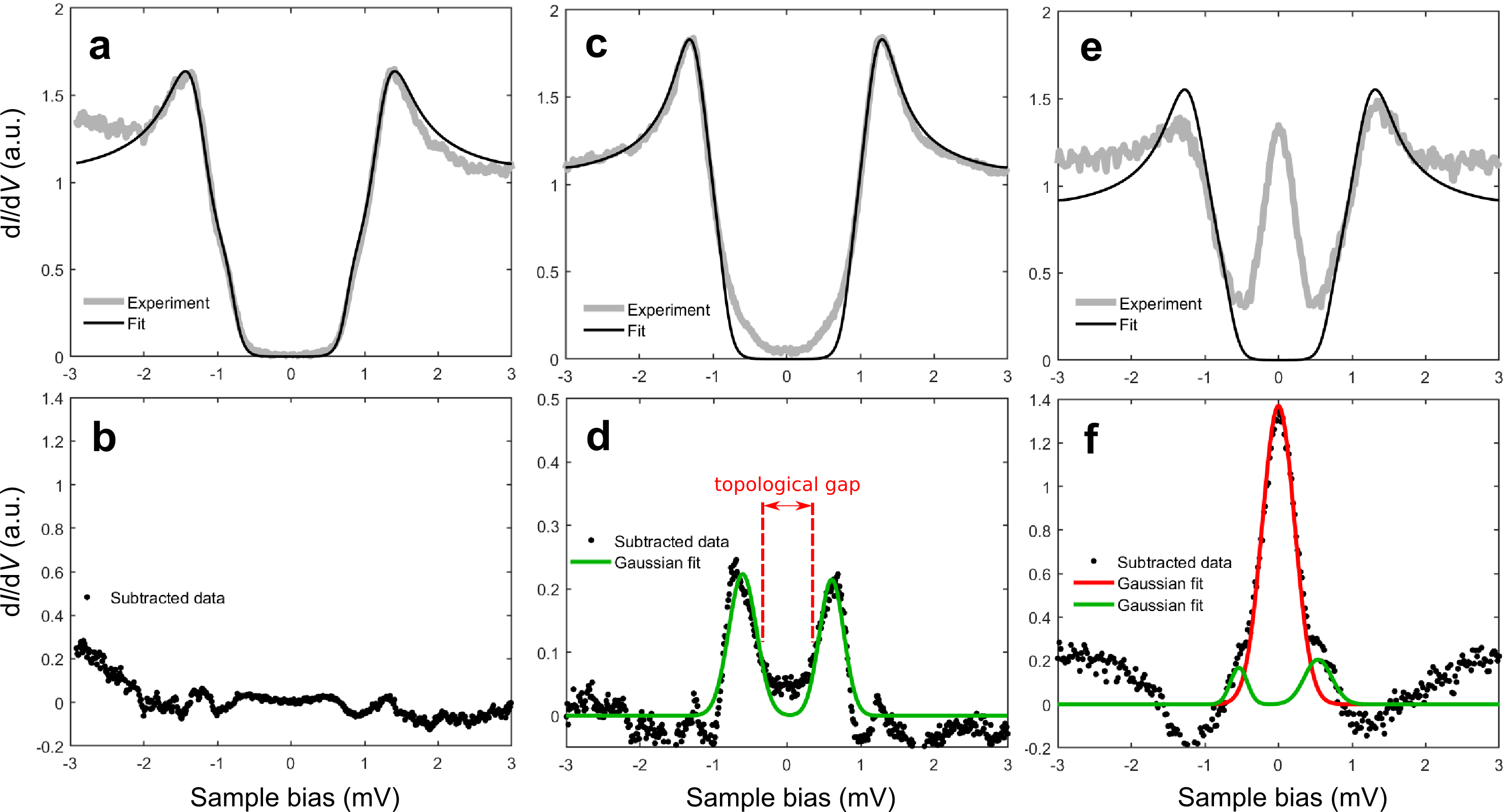}
    \caption{\textbf{Fitting the bulk NbSe$_2$ superconducting gap to a two-band model \cite{Noat2015_nbse2}.} \textbf{a,c,e}, The tunneling spectra on NbSe$_2$ (a), and in the middle (b) and at the edge (c) of an CrBr$_3$ island fitted with the two-band model. \textbf{b,d,f}, The corresponding spectra after subtracting the background given by the two-band model fit. Spectra in middle (d) and at the edge (f) of an CrBr$_3$ island show two and three-peak features within the superconducting gap, respectively. They are fitted by two or three Gaussians (green and red lines).}
    \label{ext-fig:fitting}
\end{figure}

\begin{figure}[!h]
	\centering
		\includegraphics[width=0.9\textwidth]{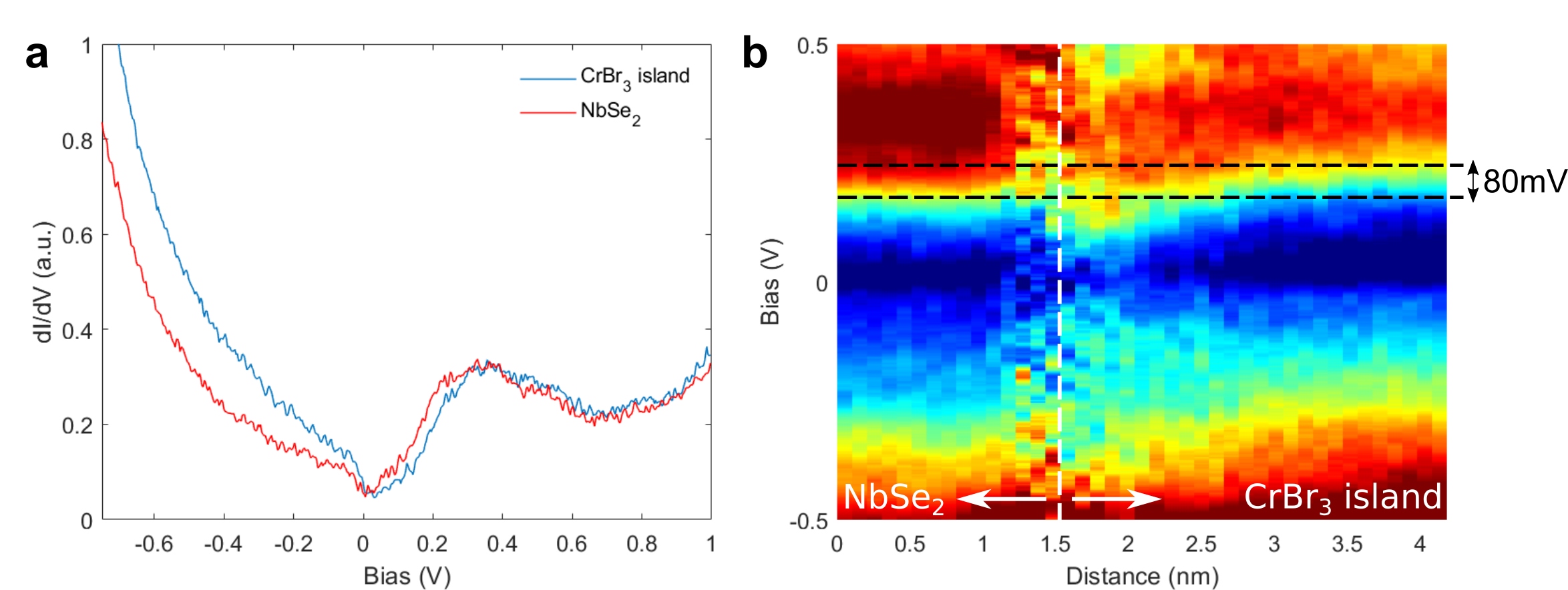}
	\caption{\textbf{Shift of the NbSe$_2$ states under CrBr$_3$. a,} Spectroscopy within the gap of the CrBr$_3$ on NbSe$_2$ (red line) and  CrBr$_3$ (blue line) showing a clear shift of the Nb d-band upwards under CrBr$_3$. \textbf{b,} Spectra recorded along a line from NbSe$_2$ to CrBr$_3$ showing the shift of ca.~80 meV. }
	\label{ext-fig:shift}
\end{figure}

\begin{figure}[!h]
	\centering
		\includegraphics[width=0.9\textwidth]{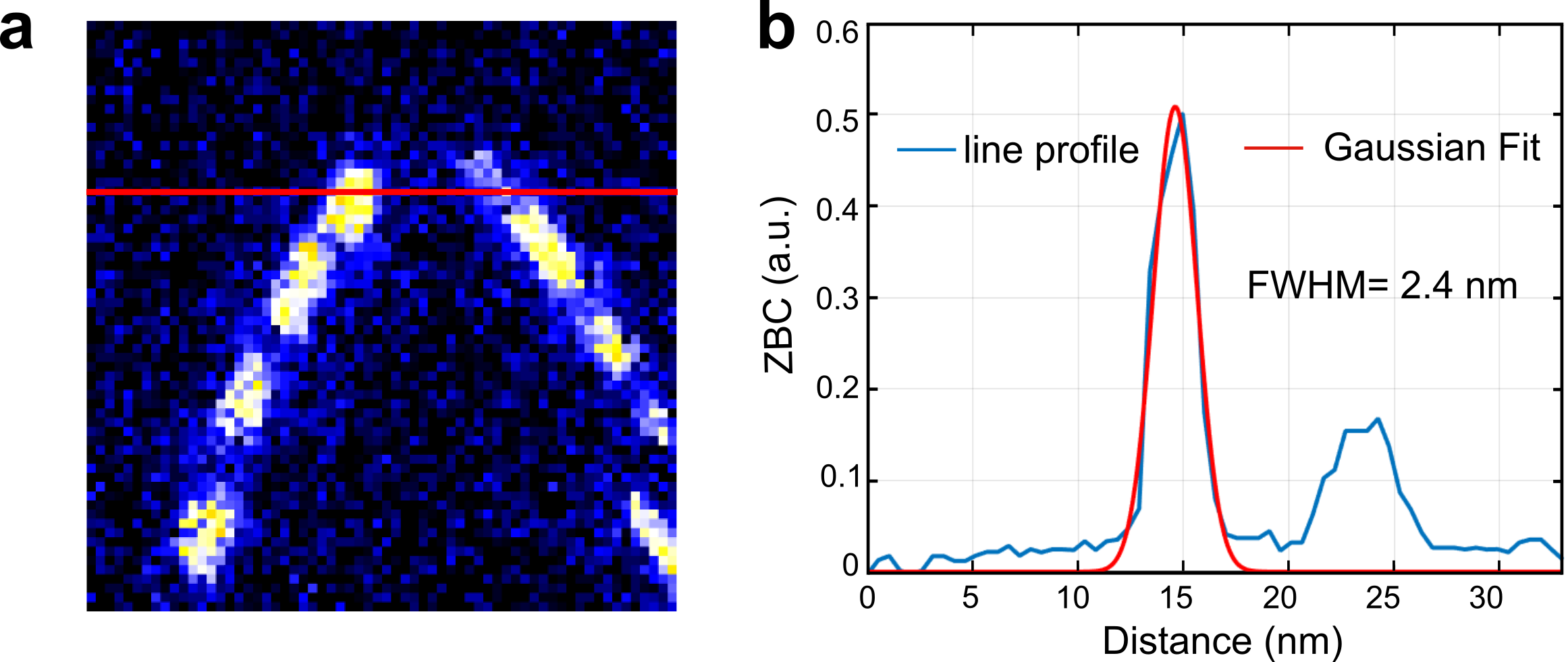}
	\caption{\textbf{Estimating the decay length of the edge states. a,} Experimentally measured differential tunneling conductance maps at zero bias. \textbf{b,} The corresponding density of states profiles across the edge of the island and corresponding Gaussian fit.}
	\label{ext-fig:edge-decay}
\end{figure}

\begin{figure}[!h]
	\centering
		\includegraphics[width=0.51\textwidth]{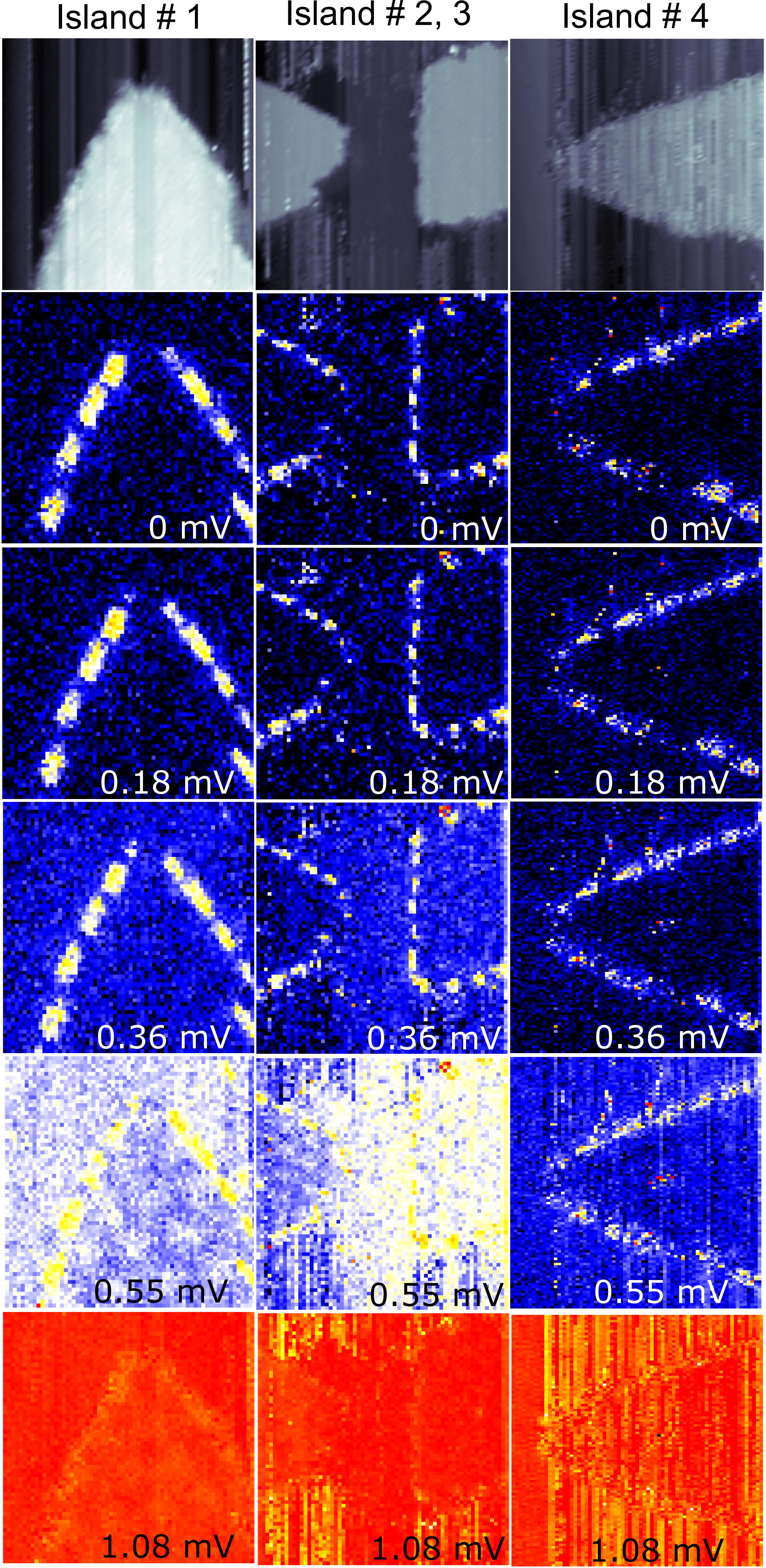}
	\caption{\textbf{Robustness and reproducibility of the edge mode.} Topography (top row, extracted during grid spectroscopy) and d$I$/d$V$ maps at different bias voltages (indicated in the different panels) on four different CrBr$_3$ islands and recorded with different microscopic tip apices. Edge modes are observed in our hybrid vdW heterostructures on all CrBr$_3$ islands, irrespective of their size and shape, or different microscopic tip.}
	\label{ext-fig:examples}
\end{figure}

\begin{figure}[!h]
	\centering
		\includegraphics[width=0.6\textwidth]{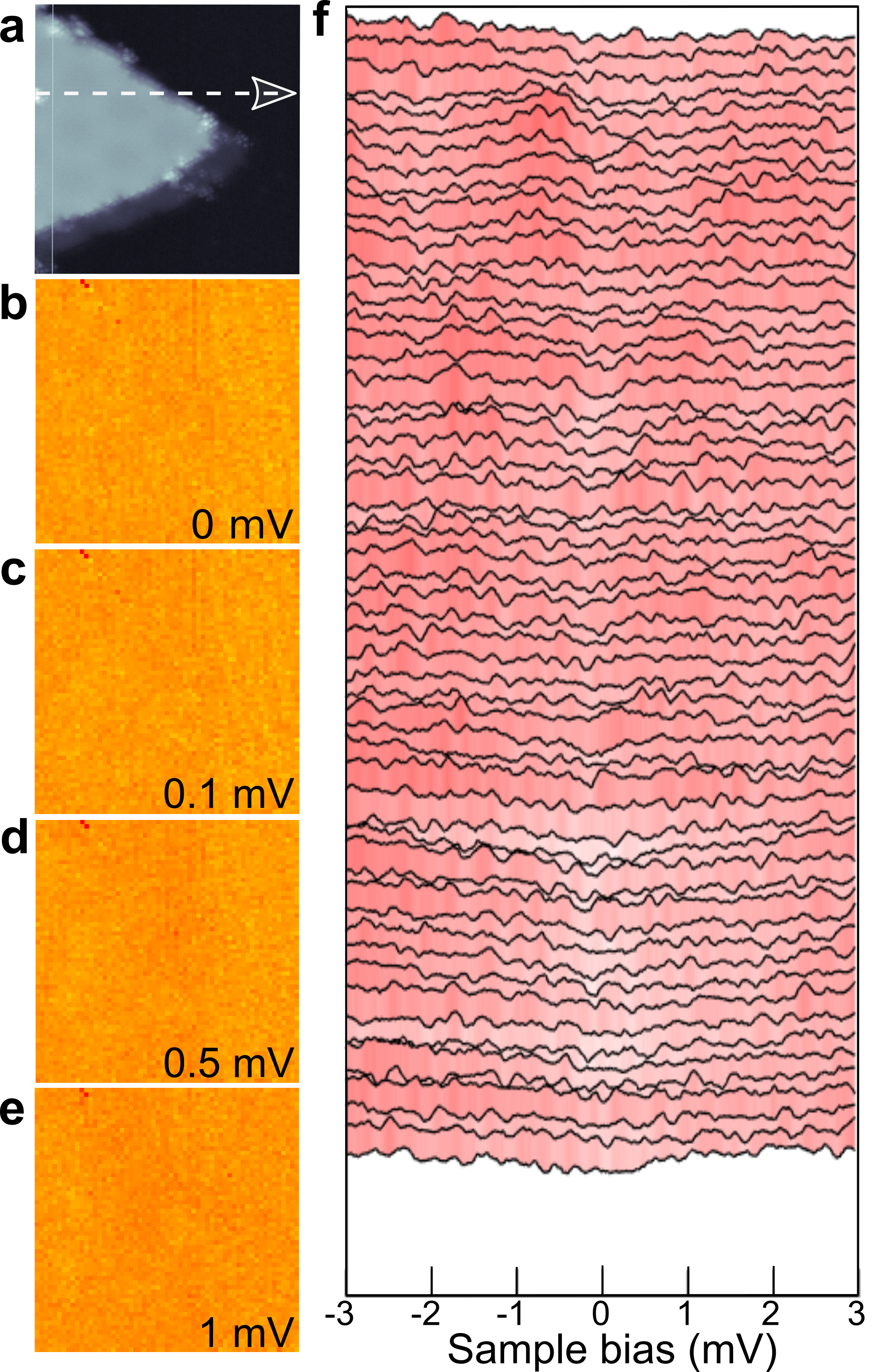}
	\caption{\textbf{Measurements in a magnetic field. a-e,} Experimentally measured differential tunneling conductance maps as a function of the bias under a 4 Tesla external magnetic field. \textbf{f,} Corresponding line spectra measured along the line indicated in panel a. STM feedback parameters: $V_\mathrm{bias} = +1$ V, $I = 10$ pA, image size: $40 \times 40$ nm$^2$. Zero bias peaks (ZBPs) can occur due to the formation of a Kondo resonance that appears when many body interactions screen magnetic impurities in metals. If the ZBPs reported here do not originate from topological origin but from the Kondo effect, they would persist beyond the superconductor-normal metal transition. However in our measurements, as soon as superconductivity is suppressed, all the states, including zero bias peaks at the edge of the island disappear and the spectrum becomes featureless.  }
	\label{ext-fig:mag-field}
\end{figure}

\begin{figure}[!h]
	\center
	\includegraphics[width=0.95\linewidth]{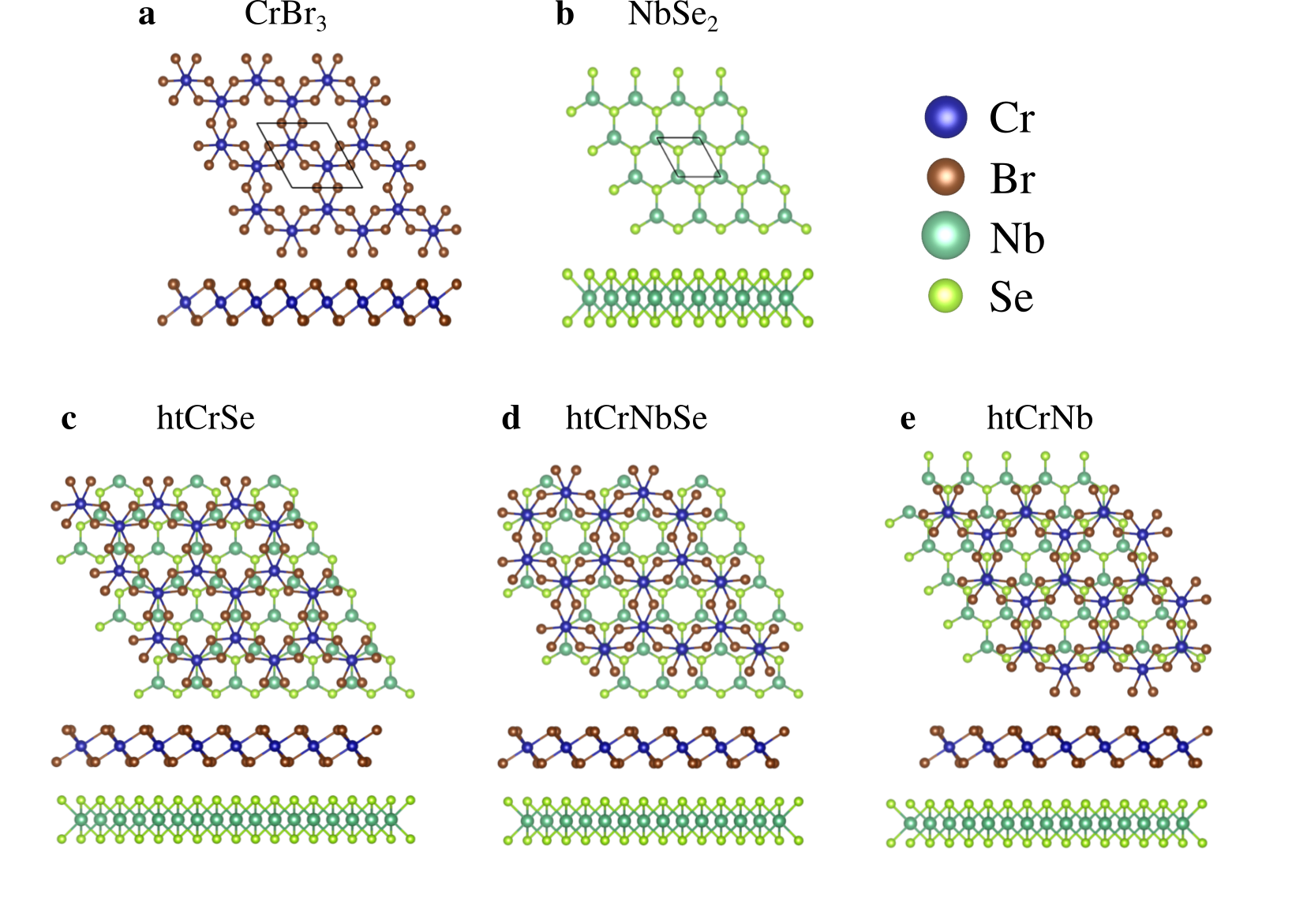}
	\caption{\textbf{Structures considered in DFT calculations. a-e,} Top and side views of the isolated CrBr$_3$ (\textbf{a}) and NbSe$_2$ (\textbf{b}) monolayers, as well as the htCrSe (\textbf{c}), htCrNbSe (\textbf{d}) and htCrNb (\textbf{e}) heterostructures. }
	\label{geos}
\end{figure}

\begin{figure}[h]
	\center
	\includegraphics[width=0.9\linewidth]{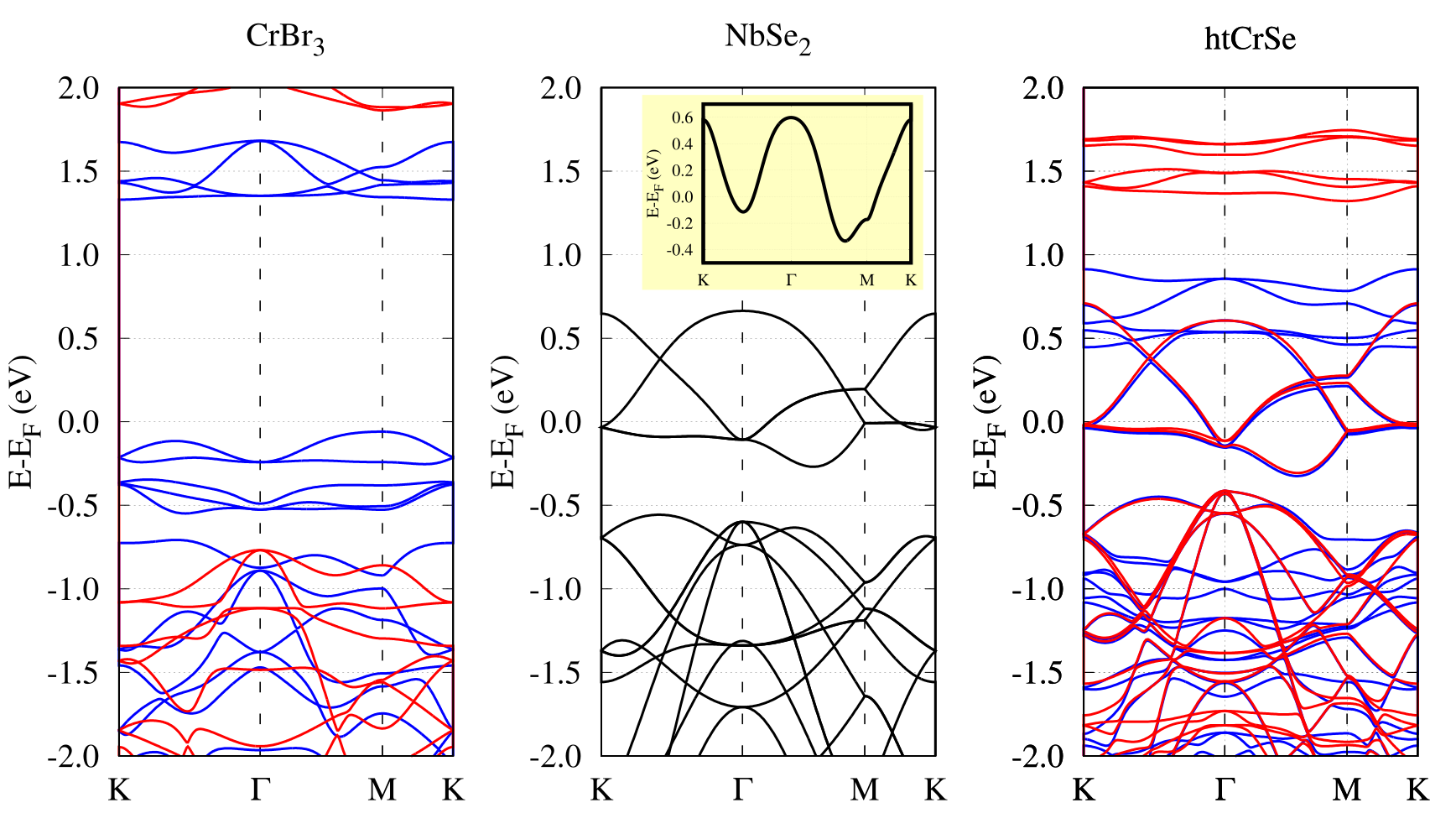}
	\caption{Band structures of the isolated monolayers CrBr$_3$ and NbSe$_2$, as well as the most stable CrBr$_3$-NbSe$_2$ heterostructure htCrSe. Blue and red  lines represent spin up and down, respectively.}
	\label{bands-ht}
\end{figure}

\begin{figure}[h]
	\center
	\includegraphics[width=0.75\linewidth]{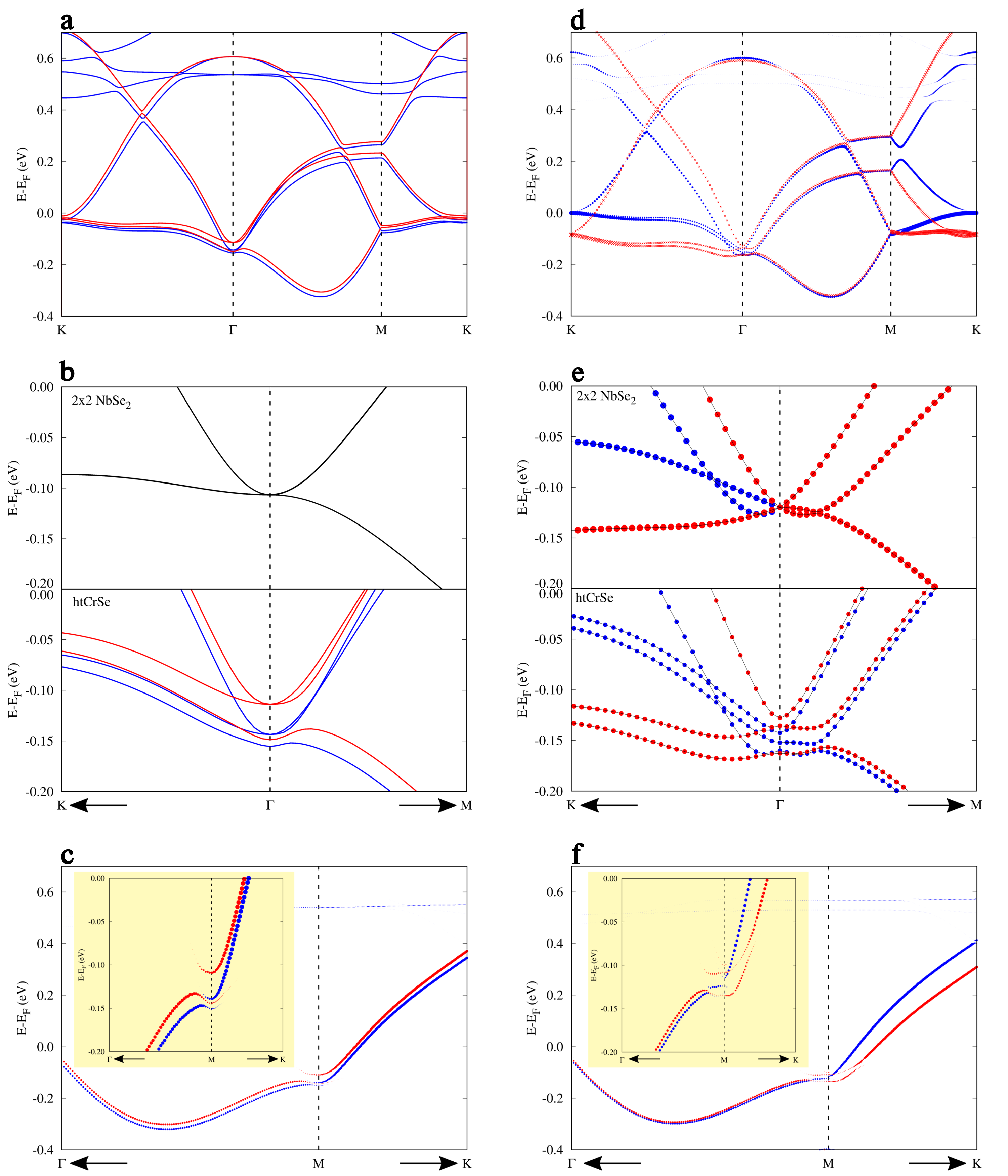}
	\caption{\textbf{a,} Spin polarized band structure of the htCrSe, where the blue and lines indicate spin up and down, respectively. \textbf{b,} Comparison between the bands of $2\times 2$ NbSe$_2$ and htCrSe near the $\Gamma$ point. \textbf{c,} Unfolded spin polarized bands. \textbf{d,} Band structure of htCrSe with SOC, where the blue (red) circles indicate the positive (negative) projection of the Nb electrons' spin on the quantization axis. \textbf{e,} Comparison between the bands with SOC of the $2\times 2$ NbSe$_2$ and htCrSe near the $\Gamma$ point. \textbf{f,} Unfolded bands obtained with SOC.}
	\label{unfold}
\end{figure}

\clearpage

\begin{table}
	\caption{Lattice parameter $L$, layer-layer distance $d$ (measured from the top Se atoms to the bottom Br atoms), and binding energy $E_b = E_{ht}- (E_\mathrm{CrBr_3} +E_\mathrm{NbSe_2})$}
	\label{tabledft}
	\begin{center}
		\begin{tabular}{ c c c c }
			\hline
			& $L$ (\AA) & $d$ (\AA) & $E_b$ (meV) \\ 
			\hline 
			htCrSe   & 6.774 & 3.275 & -284.41 \\
			htCrNbSe & 6.780 & 3.255 & -281.05  \\
			htCrNb   & 6.775 & 3.501 & -183.14 \\
			\hline
		\end{tabular}
	\end{center}
\end{table}

\end{document}